# Altermagnetic variants in thin films of Mn$_5$Si$_3$

Javier Rial,[1,*] Miina Leiviskä,[1,2] Gregor Skobjin,[3] Antonín Bad'ura,[2,4] Gilles Gaudin,[1] Florian Disdier[1], Richard Schlitz,[3] Ismaïla Kounta,[5] Sebastian Beckert,[6] Dominik Kriegner,[2] Andy Thomas,[6,7] Eva Schmoranzerová,[4] Libor Šmejkal,[2,8] Jairo Sinova,[8,9] Tomáš Jungwirth,[2,10] Lisa Michez,[5] Helena Reichlová,[2] Sebastian T. B. Goennenwein,[3] Olena Gomonay,[8,†] and Vincent Baltz[1,‡]

[1]*Univ. Grenoble Alpes, CNRS, CEA, Grenoble INP, IRIG-SPINTEC, F-38000 Grenoble, France*
[2]*Institute of Physics, Czech Academy of Sciences, Prague, Czechia*
[3]*Department of Physics, University of Konstanz, Konstanz, Germany*
[4]*Faculty of Mathematics and Physics, Charles University, Prague, Czechia*
[5]*Aix-Marseille University, CNRS, CINaM, Marseille, France*
[6]*Institute of Solid State and Materials Physics, TU Dresden, Dresden, Germany*
[7]*Leibniz Institute of Solid State and Materials Science (IFW Dresden), 01069 Dresden, Germany*
[8]*Institute for Physics, Johannes Gutenberg University Mainz, Mainz, Germany*
[9]*Department of Physics, Texas A&M University, College Station, Texas, USA*
[10]*School of Physics and Astronomy, University of Nottingham, Nottingham, United Kingdom*
[*]*javier.rialrodriguez@cea.fr; [†]ogomonay@uni-mainz.de; [‡]vincent.baltz@cea.fr*

**Abstract**

The altermagnet candidate $Mn_5Si_3$ has attracted wide attention in the context of non-relativistic spin physics, due to its composition of light elements. However, the presumed structure of the altermagnetic phase had yet to be demonstrated. In this study, we demonstrate a hallmark of altermagnetism in $Mn_5Si_3$ thin films, namely the three options, or variants, for the checkerboard distribution of the magnetic Mn atoms. The magnetic symmetries were altered by field-rotation of the Néel vector along relevant crystal directions, resulting in anomalous Hall effect anisotropy. The experimental results in nanoscale devices were corroborated by a theoretical model involving atomic site dependent anisotropy and bulk Dzyaloshinskii-Moriya interaction for a single variant.

Altermagnetism has attracted considerable attention, opening up new research paths in several fields of condensed matter physics, including spintronics, valleytronics and superconductivity, along with expectations for the emergence of closely related technological applications [1–6]. Altermagnetism features an unconventional combination of favorable characteristics of ferromagnetism and antiferromagnetism[7,8]. More specifically, an altermagnet is composed of magnetic sublattices alternately oriented in opposite directions, much like a conventional collinear antiferromagnet. It therefore presents vanishing net magnetization. It can nevertheless feature spin-current effects much like a ferromagnet [9–30]. These originate from the breaking of time-reversal ($\mathcal{T}$) symmetry, through alternate spin-splitting of the energy bands, due to non-relativistic spin-physics [31–40]. $\mathcal{T}$-symmetry breaking is also naturally manifested by antisymmetric, off-diagonal terms in physical tensors such as electrical and thermal conductivity. This yields core spintronic properties [9–30] including the anomalous Hall [9,10,14,21,24–29] and Nernst [11–13] effects. The key factor is that, unlike a conventional antiferromagnet, the magnetic sublattices that make up the altermagnet do not have the same environment, linking them by rotational symmetry rather than translational or inversion symmetry [1–6]. More generally, seen through the prism of symmetry, crystals of ferromagnets (and ferrimagnets), antiferromagnets and altermagnets form three distinct groups.

Among the altermagnetic candidates which have emerged [2,3,25,31,33,41], $Mn_5Si_3$ is composed of light-elements with weak relativistic spin-orbit coupling, making it possible to unequivocally link its altermagnetic character to non-relativistic spin physics [12–14,25–29]. Theoretical predictions [25], alongside the strong spontaneous anomalous Hall [25–27] and Nernst [12,13] effects with clear influence of crystallinity and Mn-content [13,29] despite a weak net magnetization, advocates the presence of a d-wave altermagnetic phase in this material. This phase displays a hexagonal unit cell [25,29]. It is predicted that four Mn atoms on the six possible Mn2 sites exhibit alternating magnetic dipolar ordering, resulting in a checkerboard arrangement of opposite-spin sublattices [Figure 1(a)]. As there are six possible Mn2 sites, there are three options for the checkerboard distribution of the magnetic Mn atoms among the Mn2 sites. This results in three possible structural domains, or altermagnetic variants, rotated by 120° relative to each other [Figure 1(a)], and therefore six possible magnetic domains, when taking into account reversal of the Néel vector for each variant.

Although theory and experimental evidence advocate the likely d-wave altermagnetic phase in thin films of $Mn_5Si_3$ [12–14,25–29], the specific arrangement of magnetic moments with respect to the crystal environment, i.e. the definitive proof of altermagnetism, remains to be experimentally demonstrated. An only partially successful attempt was made, based on anisotropy of physical properties [27]. More specifically, variation of the Néel vector with an external magnetic field applied at several polar and azimuthal angles reshuffled the arrangement of the moments with respect to the crystallographic axes, leading to anisotropy in the anomalous Hall response [27]. However, the effect of crystalline symmetry was shown to average out in mesoscopic devices [27], due to the presumed simultaneous presence of all altermagnetic variants [Figure 1(a)], precluding tracing the sought-after symmetries of $Mn_5Si_3$.

In this study, we demonstrate experimentally a hallmark of altermagnetism in $Mn_5Si_3$ thin films, namely the three options, or variants, for the checkerboard distribution of the magnetically ordered Mn atoms. Our experimental data reveals the symmetries of the system, at the nanoscale, through anisotropy of the anomalous Hall effect for nanometric devices. This finding is corroborated by the results of a theoretical model describing the contributions from the Néel vector and net magnetization due to canting, and their field-dependent trajectories, which were inaccessible until now.

A 17 nm-thick epitaxial thin film of $Mn_5Si_3$ was grown by molecular beam epitaxy on a 250 µm-thick insulating (ρ > 10 kΩ cm) substrate of Si(111). The growth parameters chosen were those optimized in our previous studies [29], to obtain films with the presumed altermagnetic phase [Figure 1(a)]. A few nm-thick MnSi phase acted as a priming layer to help nucleate the $Mn_5Si_3$ phase. The film contained 96 % $Mn_5Si_3$ and 4 % MnSi, and the epitaxial relationship was Si(111)[$1\bar{1}0$]//MnSi(111)[$\bar{2}11$]//$Mn_5Si_3$(0001)[$01\bar{1}0$]. It was patterned into Hall cross structures [Figure 1(b)], using electron beam lithography and ion beam etching. The longitudinal arm was oriented along the [$01\bar{1}0$] crystalline direction and, consequently, the transverse arms were oriented along the [$2\bar{1}\bar{1}0$] direction. Unless otherwise indicated, the characteristic dimension at the intersection of the longitudinal and transverse arms were 100 nm. We will see later in the text that this dimension allowed us to probe individual altermagnetic variants [Figure 1(a)]. The current density applied in the longitudinal arm was $5 \times 10^5$ A·cm$^{-2}$. The transverse voltage $V_{xy}$ was systematically measured when an external magnetic field $H$ of 2 T, i.e. sufficient to reach saturation out-of-plane [Figure 1(c)], was swept in the ($01\bar{1}0$) and ($2\bar{1}\bar{1}0$) planes and equivalents. Field sweep on a given plane was achieved by varying the $\theta$ polar angle [Figure 1(d)], and plane change was achieved by varying the $\phi$ azimuthal angle [Figure 1(e)].

Note that the transverse resistivity $\rho_{xy}$, deduced from $V_{xy}$, contains several contributions [25,27], disentangled in a control experiment consisting of varying the applied field amplitude for a fixed direction: [0001] [Figure 1(c)]. Those contributions were: (1) the linear-in-$H$ ordinary Hall background ($\rho_{OHE}$), with a slope of ∼ 0.045 µΩ·cm·T$^{-1}$, (2) a subsequent negligible symmetric-in-$H$ ($\rho_{xy,sym} = (\rho_{xy}(H) + \rho_{xy}(-H))/2$) component, supporting the collinear spin configuration[18], which will be neglected in the rest of the manuscript, and (3) the antisymmetric-in-$H$ ($\rho_{xy,asym} = (\rho_{xy}(H) - \rho_{xy}(-H))/2$) component, relating to the anomalous Hall effect and corresponding resistivity ($\rho_{AHE}$) [Figs. S1(a,b) in Supplemental material (SM)]. For all other measurements presented in this work, the $\theta$-dependences of $\rho_{xy}$ were corrected from the expected cosine-dependence of the ordinary Hall contribution: $\rho_{AHE} = \rho_{xy} - \rho_{OHE} \cos\theta$ [Figs. S1(c,d) in SM]. Figure 1(c) shows that the $H$-dependence of $\rho_{AHE}$ is in agreement with previous works [25,27] on 10 µm-wide mesoscopic Hall crosses, in terms of both amplitude and coercivity. This finding demonstrates that patterning down to 100 nm-wide Hall crosses leaves the physical properties of the system intact. The kinks observed in $\rho_{AHE}$, near remanence, are attributed to a gradual rotation of a net magnetization, before switching, as supported theoretically by Fig. S2(b). Note that the measurement temperature chosen was 110 K, also used for all the other measurements presented in this work. At this temperature, the film exhibits the characteristics of an altermagnetic phase, i.e. spontaneous anomalous Hall effect, despite a very low net magnetization of ~ 0.05 µ$_B$/u.c [12,27] [Fig. S2(a) in SM]. Its origin due to the Dzyaloshinskii-Moriya interaction (DMI) will be described later in the text, in a theoretical model.

Angular($\theta$)-depencencies of the normalized anomalous Hall resistivity $\rho_{AHE}/\rho_{AHE,max}$ when ***H*** was rotated in different crystal planes (Figure 2) allowed us to gain insight into the arrangement of magnetic moments with respect to the crystal axes of Mn$_5$Si$_3$. Based on the presumed symmetry [Figure 1(a)], data were split into two groups, according to whether ***H*** was rotated: (1) in planes which intercept two Mn2 atoms, i.e. the $(01\bar{1}0)$ plane and its two equivalents $(10\bar{1}0)$ and $(\bar{1}100)$ [Figure 2(a,c,e)], or (2) in planes which pass between four Mn2 atoms, i.e. the $(2\bar{1}\bar{1}0)$ plane and its two equivalents $(\bar{1}2\bar{1}0)$ and $(11\bar{2}0)$ [Figure 2(b,d,f)]. For all sets of data, $\rho_{AHE}/\rho_{AHE,max}$ can be seen to be maximal in amplitude when ***H*** is applied out-of-plane (i.e. along [0001], for $\theta = 0°$), and to flip sign upon *H*-reversal. This control observation is in line with the field-sweep described in Figure 1(c) and agrees with the $\mathcal{T}$-symmetry properties of the anomalous Hall effect [25,27]. When ***H*** departs from the out-of-plane direction, all $\theta$-depencencies show a multiple step-like behavior, indicating the presence of several metastable magnetic configurations, related to minima in the energy landscape. However, the step-like behavior is anisotropic as it depends upon the plane in which ***H*** is rotated. For example, the $\theta$-dependence of $\rho_{AHE}/\rho_{AHE,max}$ is significantly different when ***H*** is rotated in the $(\bar{1}2\bar{1}0)$ plane [Figure 2(b)], as it exhibits a marked hysteresis around $\theta = 90°$ and $270°$ when ***H*** is in-plane, as opposed to a plateau for all other configurations [and possibly a tiny hysteresis in Figures 2(d) and (f), which will be commented on later in the text]. These features indicate that there seems to be 2 metastable states when ***H*** is rotated in the $(\bar{1}2\bar{1}0)$ plane, as opposed to 4 when ***H*** is rotated in all other planes. Another significant feature can be seen when ***H*** is rotated in the $(10\bar{1}0)$ plane [Figure 2(e)], in which case the plateau has a negative (positive, respectively) slope around $\theta = 90°$ (270°), in contrast to opposite slopes for all other plateaus.

The step-like anisotropic behaviors, including the marked hysteresis (2 stable states) / plateaus (4) and opposite slopes discussed above for Hall crosses 100 nm-wide can be reproduced by a theoretical model (Figure 2, semi-transparent lines). The model is based on the trajectories of the magnetic moments (Figure 3), when considering a single altermagnetic variant type (ii) [Figure 1(a)]. It considers a contribution from the Néel vector, $n$ (difference between the sublattice magnetizations) and the net magnetization, $m$ (sum). The equilibrium orientation of the magnetic vectors in the external magnetic field is determined by the minimization of the magnetic energy. This energy is introduced on the basis of symmetry considerations. Starting from the crystallographic symmetry of the paramagnetic (unordered) phase, an additional altermagnetic order parameter [42] describing the local environment of the magnetic Mn2 atoms is introduced (Fig. S6 and corresponding text in SM). A DMI-like contribution $H_{\text{DMI}}(n_j m_l + n_l m_j)$ is considered. It originates from the combination of the local environment, formed by five Si and four Mn1 atoms, and deformation (Fig. S7) [25,29]. The subscripts $i$ and $j$ refer to the frame $e_j$, $e_l$, which matches the symmetry of any variant [Figure 1(a)] as follows: $e_j$ connects the two unordered Mn atoms in the basal plane, and $e_l$ points near the [0001] direction. We now focus on variant type (ii), for which $e_j$ is along $[\bar{1}2\bar{1}0]$ and $e_l$ points near the [0001] direction, in the $(\bar{1}2\bar{1}0)$ plane (Fig. S8 and corresponding text in SM). In addition, the local environment imposes the easy magnetic axis along the $[\bar{1}2\bar{1}0]$ direction. Thus, in the ground state, $n$ points along $[\bar{1}2\bar{1}0]$, which is a $C_{2x}$ symmetry axis for the variant type (ii), and due to the DMI, the magnetic moments are slightly canted, forming a non-zero magnetization along $e_l$, in agreement with the experimental findings [25,27]. The symmetry of the local environment also suggests that $\rho_{AHE}$ measured in the (0001) plane transforms in the same way as the out-of-plane component of magnetization $m_z$ and the component of the Néel vector $n_j$ along the easy axis. Thus, we assume that the transverse resistivity has two contributions, $\rho_{AHE}/\rho_{AHE,max} = \alpha \cos(\widehat{n, e_j}) + \beta \cos(\widehat{m, z})$, where: $\alpha$ and $\beta$ are phenomenological constants, $(\widehat{n, e_j})$ is the angle between $n$ and $[\bar{1}2\bar{1}0]$ and $(\widehat{m, z})$ is the angle between $m$ and [0001]. The best match with the experimental data in Figure 2 is obtained for $\alpha = 0.8$ and $\beta = 0.6$. The individual contributions from $n$ and $m$ to $\rho_{AHE}$ are shown in SM (Fig. S15). In the approximation of strong exchange coupling between magnetic sublattices, $m$ is a slave variable determined by $n$. Therefore, $\rho_{AHE}(H)$ and $n(H)$ are in one-to-one correspondence, which allows us to reconstruct the magnetic structure based on the Hall measurements for a single variant [Figure 1(a)], as will be explained below.

Our calculations mostly revealed three types of angular dependencies $n(H)$ at field values greater than the out-of-plane coercive field. We first consider the specific case when $H$ is rotated in the plane perpendicular to $e_j$, [Figure 3(a)], i. e. the $(\bar{1}2\bar{1}0)$ plane for variant type (ii) [Figure 2(b)]. Two states with opposite orientation of $n$, along $\pm e_j$, are non-degenerate $H$-rotation can only induce a step-like 180-switching. In this case, the slopes $(dn_j/d\theta)_{90°}$ and $(d\rho_{AHE}/d\theta)_{90°}$ go to infinity and the $\theta$-dependence near $\theta = 90°$ is step-like, in agreement with the experimental findings shown in Figure 2(b). The stability of these states is determined mainly by the magnetic anisotropy resulting in a wide hysteresis and large coercivity in the $\theta$-scans. We next consider the case when $H$ has a non-zero projection on $e_j$ [Figure 3(b)], i.e. when it is rotated in the $(10\bar{1}0)$ plane for variant type (ii) [Figure 2(e)]. This can induce a spin-flop transition leading to the trapping of $n$ in one of two metastable states close to out-of-plane: $n||\pm e_l$. In this case, $n$ hops between four states during $H$-rotation: $n||e_j \Rightarrow e_l \Rightarrow -e_j \Rightarrow -e_l$. The stability of the additional, "out-of-plane" states is determined by the DMI, which sets the hard magnetic axes along $e_j \pm e_l$ and along $e_k \perp e_j, e_l$. Thus, the energy trap for these states is maximal when $H$ is rotated in the $(e_j, e_l)$ plane, i. e. $(10\bar{1}0)$ for variant type (ii). In this case, the stability range of the out-of-plane states is large, which leads to the large hysteresis and strong coercivity [Figure 2(e)]. Moreover, the slope $(dn_j/d\theta)_{90°}$ is negative and $(d\rho_{AHE}/d\theta)_{90°}$ can be either negative or positive depending on the ratio $\alpha/\beta$ of the contributions from $n$ and $m$ [Figure 2(e), and SM, Fig. S15]. We finally consider all other intermediate cases [Figures 3(c) and 2(a,c,d,f)]. From maximum when $H$ is rotated in the $(e_j, e_l)$ plane, the stability range of out-of-plane state decreases for other planes of rotation. As a result, the slope $(dn_j/d\theta)_{90°}$ evolves continuously from a maximum negative value to large positive values for $H \perp e_j$, where the out-of-plane states are unstable. This also results in a variation of the hysteresis width, which is pronounced in two limiting cases and almost disappears for intermediate cases.

We note that the model can also reproduce $\rho_{AHE}(H)$ shown in Figure 1(a), for $H$ swept with variable amplitude along [0001] [Fig. S2(b)]. In addition, we would like to point out that the initial model of altermagnetism in Mn$_5$Si$_3$ was deduced from the four magnetically ordered Mn2 atoms, which are at Wyckoff positions that are conducive to d-wave altermagnetism with the d$x^2$-$y^2$ symmetry [25]. With this symmetry, a Néel vector pointing in-plane along an axis of high symmetry, e.g. $[2\bar{1}\bar{1}0]$ and its two equivalents, was forbidden as it did not allow an out-of-plane Hall vector. The model returned a ground state in which the Néel vector was along the $[2\bar{2}01]$ crystal direction ($[111]$ direction in the 3-component a – b – c notation). This direction was then used as an input parameter in the atomic spin simulations of Refs. 14 and 26. Our refined model showed that the symmetry of the altermagnetic phase of Mn$_5$Si$_3$ is actually lower. Here, the Néel vector was derived from energy minimization from the non-magnetic phase, considering the local environment of the Mn2 atoms. The environment corresponding to the crystallographic positions reproduces the d$x^2$-$y^2$ symmetry of the initial model. Convergence of energy minimization requires inhomogeneous displacements of Si and Mn1 atoms, which reduces the symmetry from a magnetically unordered but already altermagnetic phase down to the monoclinic phase. In this phase of lower symmetry, $n$ pointing in-plane along an axis of high symmetry allows an out-of-plane Hall vector, in contrast to the phase of higher symmetry initially considered. Our model thus returned a ground state in which the Néel vector connects the two unordered Mn atoms in the basal plane. For variant type (i), (ii), and (iii), $\boldsymbol{n}$ is along $[2\bar{1}\bar{1}0]$, $[\bar{1}2\bar{1}0]$, and $[11\bar{2}0]$, respectively. We note that, compared to our epitaxial thin films, the magnetic phase of Mn$_5$Si$_3$ in bulk and in polycrystals is not altermagnetic and displays a different orientation of $\boldsymbol{n}$ [43,44].

Based on the relationship between the altermagnetic variants [Figure 1(a)] the features corresponding to variant type (ii) (e.g. the marked hysteresis for $\phi = 30°$ around the in-plane $\boldsymbol{H}$ direction), are rotated by 120° in plane for type (iii) variant and -120° in plane for type (i). Consequently, we can conclude that the weak hysteresis observed around the in-plane $\boldsymbol{H}$ direction in Figures 2(d) and 2(f), i.e. for $\phi = 90$ and 150°, respectively, highlight contributions from the presence of minority variants of types (i) and (iii) [Figure 1(a)] in the 100 nm-wide Hall cross we measured. We assume that local defects can lift the degeneracy between the variants and locally favor one over the other. Experiments on another such Hall cross, located elsewhere in the sample and oriented in the same way, enabled us to identify a region where the altermagnetic variant type (iii) [Figure 1(a)] was in the majority (Fig. S4). This conclusion could be drawn because the magnetic field-related anisotropy pattern (set of 6 angular-dependences) was shifted by +120° compared to the pattern we observed for variant type (ii) (Fig. 2). We can also deduce that patterning does not favor a specific variant, for example through strain.

Ultimately, when all altermagnetic variants are equally populated, i.e. for a mesoscopic 10 μm-wide Hall cross, the set of data when $\boldsymbol{H}$ is rotated in the $(2\bar{1}\bar{1}0)$ plane and its two equivalents superimpose [Figures 4(a)] [27]. In fact, for such a mesa, the footprint of variant (i), (ii), and (iii), i. e. the marked hysteresis around $\theta = 90°$ and 270°, occurs when $\boldsymbol{H}$ is rotated in the $(2\bar{1}\bar{1}0)$, $(\bar{1}2\bar{1}0)$, and $(11\bar{2}0)$ plane, respectively (Fig. S18-20). Because the footprints superimpose, and because the current is always along $(01\bar{1}0)$, i. e. oriented differently for all three variants in the mesa, we can confirm[27] that the anisotropic pattern of the anomalous Hall effect relates to the crystal plane in which the field rotates, rather than to the crystal axis along which the current channel is oriented. We note that current-related effects may occur for large current densities, such as the spin-splitter effect, in which anisotropic spin polarization can take place [45]. The mesoscopic configuration is equivalent to averaging over the signals obtained for a given variant [Figures 4(b)]. Similar results were obtained for data sets when $\boldsymbol{H}$ is rotated in the $(01\bar{1}0)$ plane and its two equivalents [Figures 4(c,d)]. We note that the use of larger external fields of 4 and 7 T causes a reduction of the hysteresis in the angular scan [27] (also Figs. S16,17). The stable and metastables states, which are key to the deduction of the variant, and the conclusions of our work are unaffected.

In conclusion, the main contribution of this paper is the experimental demonstration of the, so far, presumed altermagnetic arrangement of magnetic moments with respect to crystal symmetries in thin films of $Mn_5Si_3$, via identification of variants, i.e. the three options for the checkerboard distribution of the magnetic Mn atoms [Figure 1(a)]. Our experimental data revealed the symmetries of the system, at the nanoscale, through anisotropy of the anomalous Hall effect for nanometric devices. They elevate $Mn_5Si_3$ from a candidate to a proven altermagnet. The magnetic moment-crystal axes arrangement was reshuffled via field-driven variation of the Néel vector, with an external magnetic field applied at several polar and azimuthal angles. These findings were corroborated by the results of a refined theoretical model accounting for the local environment of the magnetic sublattices. Corroborating the model and the experiments revealed the ground state, as well as the trajectories of the Néel vector and net magnetization due to DMI-induced canting. This information was inaccessible until now, because variants were not isolated. Our study therefore opens up prospects for a wide range of further fundamental research and practical applications in altermagnetism, where detecting and controlling altermagnetism is critical. We also wish to open the conversation on the implications of variants. A magnetic variant differs from a magnetic domain, in that a single variant can contain two magnetic domains of opposite direction. Magnetic variants, therefore, bring a new degree of freedom to the system. How to exploit this degree of freedom, how to control it, what is the ground state between two variants, and how do variants evolve across the Néel temperature are open questions. For example, a magnetic field is not expected to directly modify non-magnetic atoms, and the process by which, e.g. strain, one variant can be more densely populated than the other is an open challenge.

**Acknowledgments**

We thank G. Atcheson for critical reading of the manuscript. This work was supported by the French national research agency (ANR) and the Deutsche Forschungsgemeinschaft (DFG) (Project MATHEEIAS - Grant No. ANR-20-CE92-0049-01 / DFG-445976410, and Project PEPR SPIN - Grant No. ANR-22-EXSP-0007). H. R., E. S., S. T. B. G. and A. T. are supported by the DFG-GACR grant no. 490730630. J. R. acknowledges MINECO for the Margaritas Salas program. D. K. acknowledges the academy of sciences of the Czech Republic (Project No. LQ100102201). O. G. and J. S. acknowledge funding by the DFG Grant No. TRR 288-422213477 (projects A09 and A12) and TRR 173-268565370 (projects A03, A11, and B15).

**Figures and captions**

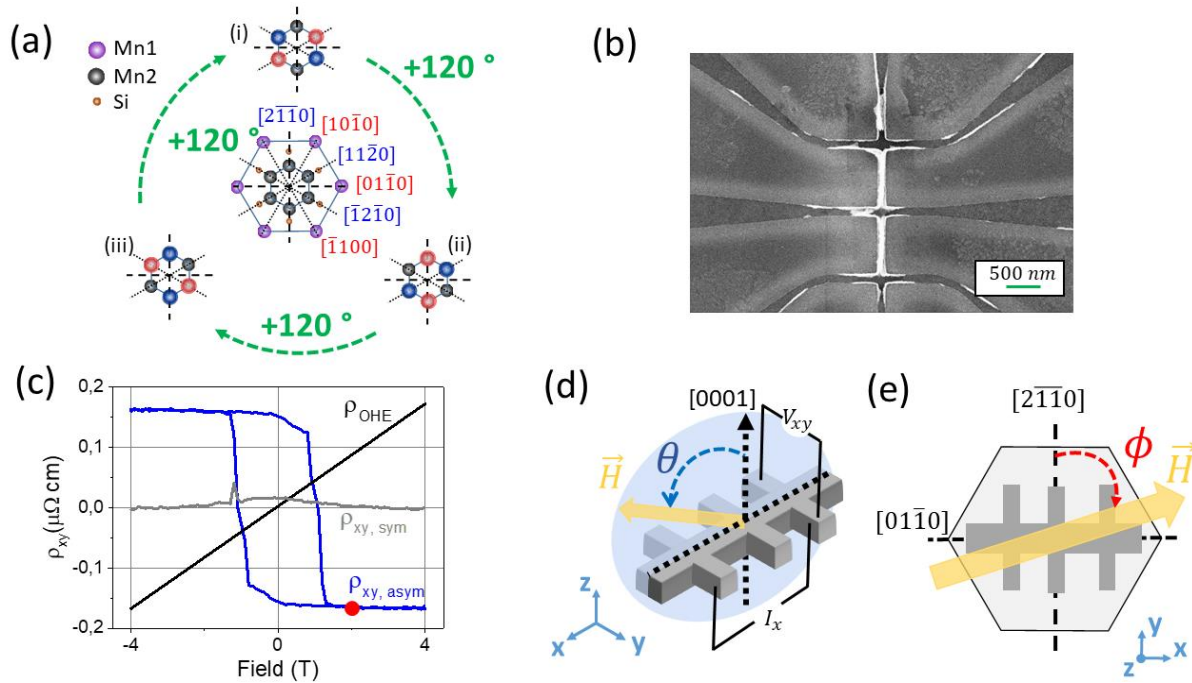

Figure 1. (Color Online) (a) Representation of the hexagonal unit cell of the altermagnetic phase of $Mn_5Si_3$, consisting of two magnetically ordered opposite-spin sublattices of Mn2-atoms, highlighted in red and blue. The 3 possible variants rotated by 120° are shown. (b) Scanning electron microscopy image of the ~100 nm-wide Hall cross structure used in this work. The white highlighted foil is mostly the negative resist MaN2403, which was burned during ion etching. Re-deposition only around the corners of the Hall bar's arms cannot be excluded, due to the small size of the device. See also Fig. S5. (c) The 3 contributions to the field (*H*)-dependence of the transverse resistivity, $\rho_{xy}$, measured at *T* = 110 K, for *H* along [0001]. (d,e) Representations of the experimental design, highlighting the $\theta$ polar angle used for field-sweep on a given plane. Plane change was achieved by varying the $\phi$ azimuthal angle.

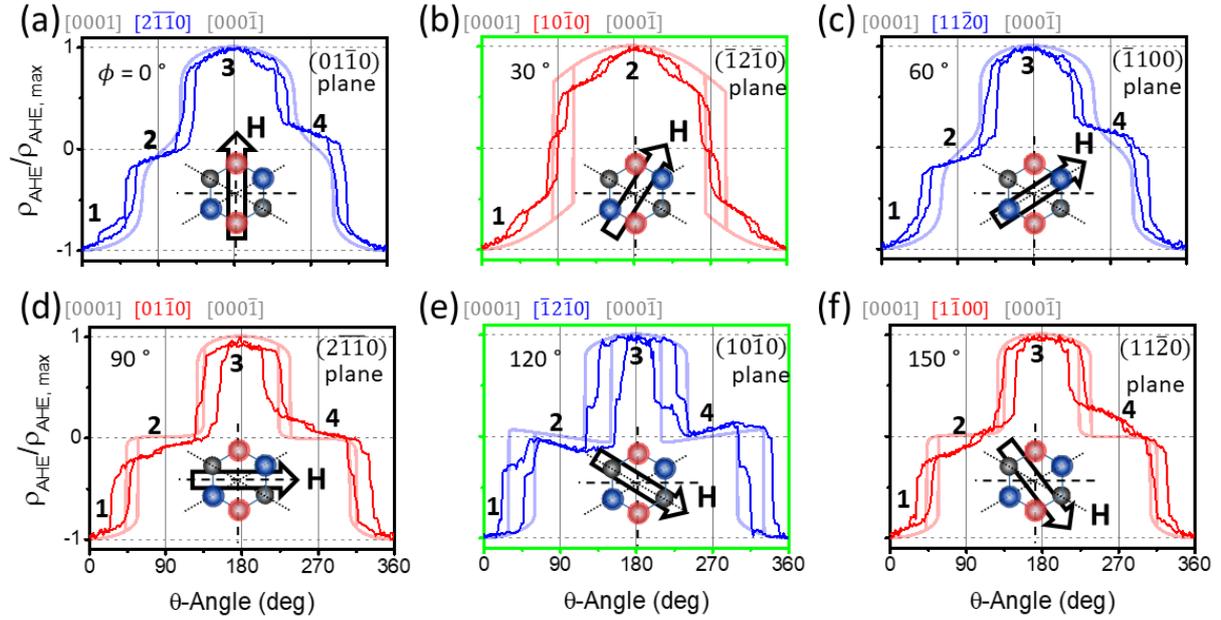

Figure 2. (Color Online) Angular($\theta$)-depencence of the normalized anomalous Hall resistivity $\rho_{AHE}/\rho_{AHE,max}$ measured for a ~100 nm-wide Hall cross, when *H* of 2 T is rotated in different crystal planes (Full lines: experiments and semi-transparent lines: theory). (a,c,e) For $\phi = 0, 60, 120°$, *H* is rotated in the $(01\bar{1}0)$ plane and equivalent. (b,d,f) For $\phi = 30, 90, 150°$, *H* is rotated in the $(2\bar{1}\bar{1}0)$ plane and equivalent. The majority altermagnetic variant studied here is type (ii) [see Figure 1(a)], which we deduced by correlating experiments and theory.

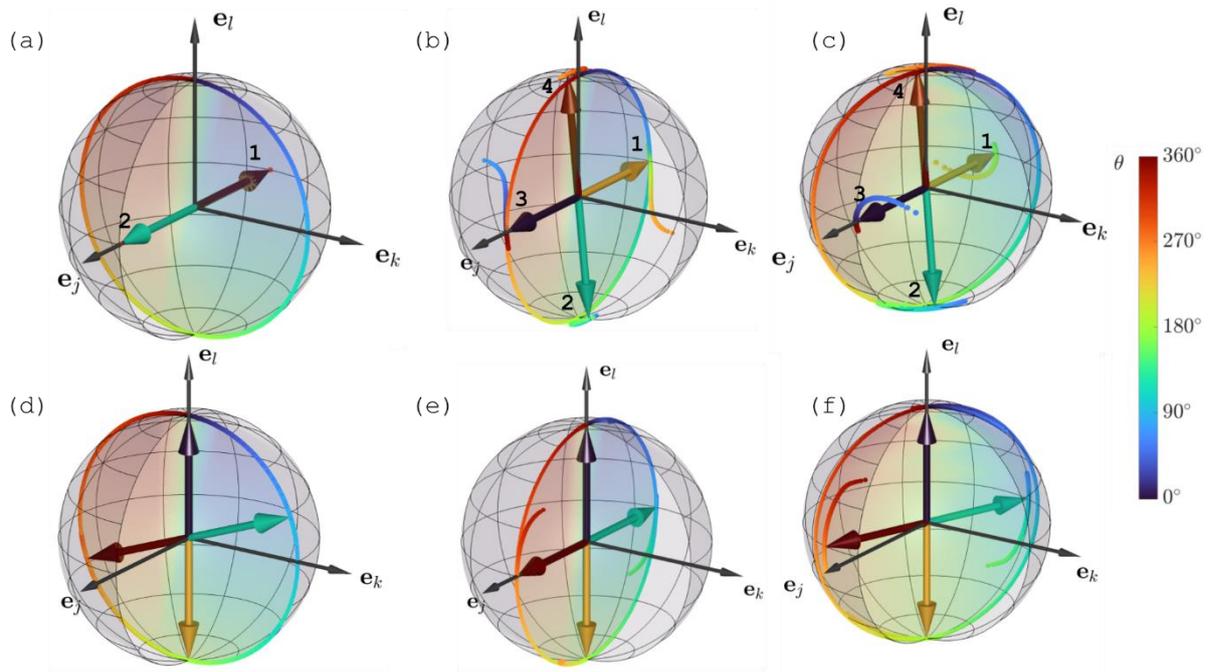

Figure 3. (Color Online) (a,b,c) Calculated states of the Néel vector, *n*, and (d,e,f) net magnetization, *m*, when *H*, larger than the coercive field, is rotated in three different planes. The frame $e_j, e_k, e_l$ depends on the altermagnetic variant considered [Figure 1(a)]. For example, for variant (ii), (a,d) corresponds to the case when *H* is rotated in the $(\bar{1}2\bar{1}0)$ plane [Figure 2(b)], (b,e) in the $(10\bar{1}0)$ plane [Figure 2(e)], and (c,f) in any other plane [Figure 2(a,c,d,f)]. The outline of the solid disk describes the trajectory of *H*. The arrows describe the orientations of *n* and *m*, in panels (a,b,c) and (d,e,f), respectively for selected orientations of *H*. The color code indicates the correspondence between the orientations of *n* (*m*) and *H*. Discontinuous traces are linked to transient states, when *n* (*m*) switches from one state to another. The numbers (1,2) and (1,2,3,4) correspond to those in Figure 2.

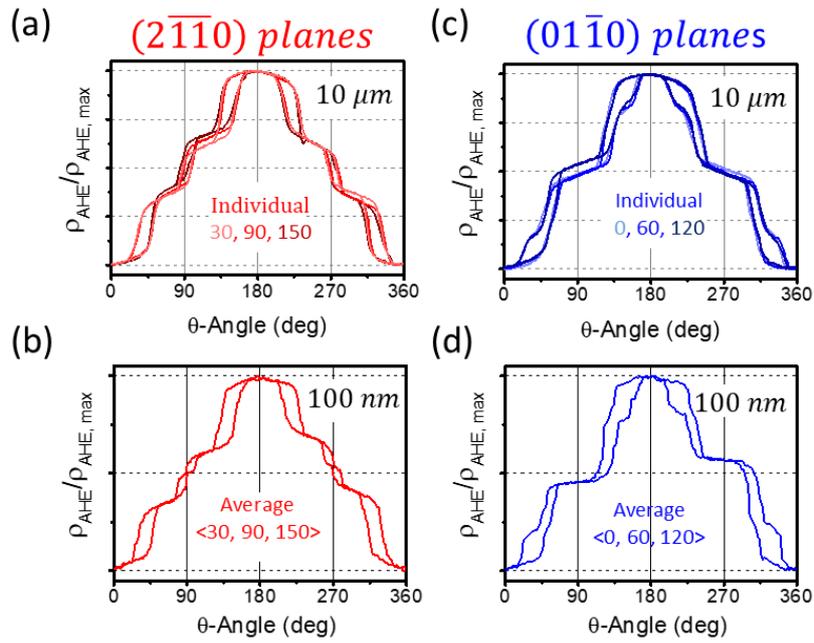

Figure 4. (Color Online) (a,c) Superposition of the individual Hall signals, when *H* is rotated in the $(2\bar{1}\bar{1}0)$ plane and its two equivalents [$(01\bar{1}0)$, respectively], measured for a 10 μm-wide Hall bar containing all three variants [Figure 1(a)]. (b,d) Averaged Hall response over all signals obtained in Figure 2(b,d,f) [(a,c,e), respectively], when *H* is rotated in the $(2\bar{1}\bar{1}0)$ plane and equivalents [$(01\bar{1}0)$, respectively], i.e. for $\phi = 0, 60, 120°$ ($\phi = 30, 90, 150°$, respectively), for a 100 nm-wide Hall cross containing mainly one altermagnetic variant (see Figure 2 for the individual signals).